\documentclass[11pt,a4paper]{article}
\pdfoutput=1
\usepackage[utf8]{inputenc}
\usepackage[T1]{fontenc}
\usepackage{lmodern}
\pdfmapfile{+lm.map}
\usepackage{amsmath,amssymb,mathtools}
\usepackage[margin=1in]{geometry}
\usepackage{microtype}
\usepackage{booktabs,longtable,array,calc}
\usepackage{caption}
\usepackage{float}
\usepackage{graphicx}
\usepackage{xcolor}
\usepackage{listings}
\usepackage{newunicodechar}
\usepackage{hyperref}
\usepackage{bookmark}
\usepackage{xurl}
\hypersetup{colorlinks=true,linkcolor=blue!55!black,urlcolor=blue!55!black,pdftitle={Negative and Fractional Types in the Fidelity Framework: Computation, Reversibility and Resource Pairing},pdfauthor={Houston Haynes}}
\newcommand{\pandocbounded}[1]{\begingroup\setbox0\hbox{#1}\ifdim\wd0>\linewidth\resizebox{\linewidth}{!}{#1}\else#1\fi\endgroup}
\newsavebox{\paperdisplaybox}
\let\paperdisplayopen\[
\let\paperdisplayclose\]
\long\def\[#1\]{\begingroup
  \sbox{\paperdisplaybox}{$\displaystyle #1$}%
  \ifdim\wd\paperdisplaybox>1.25\linewidth
    \PackageError{paper}{Display requires source line breaks}{Split the display in Markdown.}%
  \fi
  \paperdisplayopen
  \ifdim\wd\paperdisplaybox>\linewidth
    \typeout{Paper display scaled to fit width: \the\wd\paperdisplaybox}%
    \resizebox{\linewidth}{!}{\usebox{\paperdisplaybox}}%
  \else\usebox{\paperdisplaybox}\fi
  \paperdisplayclose\endgroup}
\newunicodechar{α}{\ensuremath{\alpha}}
\newunicodechar{β}{\ensuremath{\beta}}
\newunicodechar{ε}{\ensuremath{\varepsilon}}
\newunicodechar{θ}{\ensuremath{\theta}}
\newunicodechar{κ}{\ensuremath{\kappa}}
\newunicodechar{Σ}{\ensuremath{\Sigma}}
\newunicodechar{⊗}{\ensuremath{\otimes}}
\newunicodechar{⊕}{\ensuremath{\oplus}}
\newunicodechar{↔}{\ensuremath{\leftrightarrow}}
\newunicodechar{¬}{\ensuremath{\neg}}
\newunicodechar{∧}{\ensuremath{\land}}
\newunicodechar{≤}{\ensuremath{\leq}}
\newunicodechar{×}{\ensuremath{\times}}
\newunicodechar{·}{\ensuremath{\cdot}}
\newunicodechar{†}{\ensuremath{\dagger}}
\newunicodechar{⟩}{\ensuremath{\rangle}}
\newunicodechar{²}{\ensuremath{{}^2}}
\newunicodechar{¹}{\ensuremath{{}^1}}
\newunicodechar{⁻}{\ensuremath{{}^-}}
\newunicodechar{ᵢ}{\ensuremath{{}_i}}
\newunicodechar{₀}{\ensuremath{{}_0}}
\newunicodechar{₁}{\ensuremath{{}_1}}
\newunicodechar{₂}{\ensuremath{{}_2}}
\newunicodechar{₊}{\ensuremath{{}_+}}
\begin{document}
\begin{center}

\textbf{Negative and Fractional Types in the Fidelity Framework}

\emph{Computation, Reversibility and Resource Pairing}

\textbf{Houston Haynes}\\
Braidpoint, Asheville, NC\\
\href{mailto:research@braidpoint.tech}{\nolinkurl{research@braidpoint.tech}}\\
June 2026

\end{center}

\section*{Abstract}\label{abstract}
\addcontentsline{toc}{section}{Abstract}

Programs routinely create temporary data, retain state and revisit earlier results. Negative and fractional types benefit those activities in a type-resolved account of computation. Negative values redirect evaluation. Fractional values introduce value-specific auxiliary data and discharge it after a computation restores the required value. We develop both as first-class computational forms in our Clef design, using Chen and Sabry\textquotesingle s directed, value-indexed calculus and distinguishing James and Sabry\textquotesingle s logic-variable search interpretation.

Our Program Semantic Graph retains their value identities and resource uses, enabling Composer to preserve their behavior through MLIR operations and native realization. In this paper, a negative-type counter and a fractional workspace calculation demonstrate the two constructions. Local proofs establish preservation of matching and independent resource discharge. Dimensional inference and proof terms supply premises through our four verification tiers.

The applications connect deferred reverse computation to gradient calculations without an activation tape, and checked inverse steps to in-memory event recovery. A finite reconstruction horizon governs the exchange between storage and recomputation. We derive its error recurrence and show where Urschel\textquotesingle s elimination-growth bounds can support a realized inverse step. Bayesian updates and quantum circuits supply further computational cases, including temporary workspace, measurement and a worked adiabatic schedule. Exact matching, bounded reconstruction and physical realization retain their respective contracts.

The resulting architecture makes the information needed to run, reverse or complete a computation available to the compiler. Libraries can expose these operations through familiar interfaces while their retained structure supports unboxed execution and differential compilation across targets.

\section*{1. Computation and Retained Information}\label{computation-and-retained-information}
\addcontentsline{toc}{section}{1. Computation and Retained Information}

A program often keeps an intermediate value because later work will need it. A gradient calculation revisits earlier activations. A service retains state so an update can be reversed. A temporary workspace must reach an allowed final state before its storage can be reused. These decisions connect the meaning of a computation to its memory requirements, yet ordinary function signatures describe little of that connection.

Our interest in negative and fractional types began with machine learning and quantum computation. Deferred reverse steps suggested a way to recover values when a gradient needs them, while quantum circuits made the creation and cleanup of auxiliary state explicit. In-memory event recovery presents a more familiar version of the same engineering concern. An application should be able to reconstruct a state, or complete the use of a temporary resource, with a contract precise enough for the compiler to preserve and optimize.

Negative and fractional types bring these recovery and resource-use relationships into the program\textquotesingle s type structure. Managing the relationships through separate logs, flags and cleanup paths spreads an algorithm\textquotesingle s invariants across bookkeeping code. This enlarges the implementation and makes correctness depend on agreement among those paths. A compositional type structure makes the dependencies explicit and connects each operation to the premises that justify its use.

We take the operational meaning of these constructions from the calculi developed by James and Sabry and subsequently Chen and Sabry. A negative value changes the direction of evaluation. A fractional value participates in the creation and elimination of auxiliary information \hyperref[ref-james-dualities]{{[}1{]}} \hyperref[ref-chen]{{[}2{]}}. The two constructions have executable rules. Their types determine how they compose, and their proofs establish which equations those compositions satisfy.

We develop a Clef interpretation in which both constructions are first-class forms and their proof dependencies remain available throughout compilation. A program can pass their values through typed interfaces, combine their operations and defer a computation that uses them. Resource-bearing values retain their use restrictions through those steps. Where the retained evidence justifies reconstruction or reuse, Composer can remove redundant state and bookkeeping while preserving the algorithm\textquotesingle s contract. The examples below begin with a counter that routes through a negative alternative and a calculation that uses a fractional value to complete a temporary workspace\textquotesingle s lifetime.

\subsection*{1.1 Contribution and Compiler Context}\label{contribution-and-compiler-context}
\addcontentsline{toc}{subsection}{1.1 Contribution and Compiler Context}

Our contribution is an integration of these calculi with native compilation and automatic verification. We specify the value and instance identities that our graph retains, give local preservation results, and connect checked reversal to a quantitative choice between retained state and recomputation. The applications derive concrete consequences under stated premises. The calculus and its Clef interfaces form a language design, with library operations used below as illustrative source notation.

Our Native Type Universe (NTU) keeps source identity available when a target representation is selected. DTS/DMM develops its dimensional and memory-lifetime account, so an implementation can retain a quantity\textquotesingle s meaning while choosing its storage \hyperref[ref-dts]{{[}3{]}}. PHG supplies the joint relationships needed to connect an operation with all the values on which its validity depends \hyperref[ref-phg]{{[}4{]}}. These relationships are particularly useful for a fractional operation, whose input carrier, value index and resource use must agree at one site.

The verification architecture in DBC connects local inference to reusable proofs, including concurrent and relational reasoning \hyperref[ref-dbc]{{[}5{]}}. FPS develops preservation through our thin MLIR middle end and the dependencies that support differential compilation \hyperref[ref-fps]{{[}6{]}}. Here we supply the operational content those mechanisms preserve for negative and fractional programs. ADM provides the companion account of adaptive computation and arithmetic selection used in the learning applications \hyperref[ref-adm]{{[}7{]}}.

\section*{2. Two Computational Dualities}\label{two-computational-dualities}
\addcontentsline{toc}{section}{2. Two Computational Dualities}

Reversible programming begins with operations that retain enough information to retrace their action. Products describe simultaneous components of a value, and sums describe alternatives. James and Sabry extend this finite-type setting with negative and fractional constructions, exploring the information effects of introducing and eliminating data \hyperref[ref-james-information]{{[}8{]}} \hyperref[ref-james-dualities]{{[}1{]}}. Chen and Sabry give a directed abstract machine and a value-indexed account of fractional operations \hyperref[ref-chen]{{[}2{]}}.

We use Chen and Sabry\textquotesingle s operations as the reference core. Their value indices specify the behavior of creation, matching and elimination. James and Sabry\textquotesingle s logic-variable interpretation supplies a separate computational route in which unification resolves a demand. Both belong in the language-design account, with their respective evaluation rules determining which programs and transformations they admit.

\subsection*{2.1 Negative Values and Directed Evaluation}\label{negative-values-and-directed-evaluation}
\addcontentsline{toc}{subsection}{2.1 Negative Values and Directed Evaluation}

A negative value carries evaluation in the opposite direction through the reference machine. Its introduction and elimination operations provide turns in a program\textquotesingle s control flow. A composition containing two suitable turns can return the original value while evaluation has travelled through both directions.

The following ladder records that composition. Its bent path is the additive snake identity, expressed operationally through the two turns.

\begin{figure}
\centering
\pandocbounded{\includegraphics[keepaspectratio]{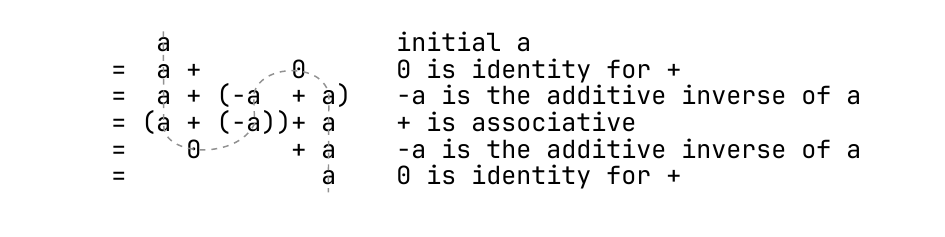}}
\caption{Additive snake construction, showing a value returning through negative-type pairing.}
\end{figure}

\emph{The sum notation describes the directed calculus. Its intermediate creation and cancellation steps have that operational interpretation. Ordinary Clef value sets retain their usual meaning, including the empty type \(0\).}

With additive unit \(0\), the signatures have the form

\[\eta_A^+ : 0\longrightarrow\operatorname{Neg}(A)+A,
\qquad
\varepsilon_A^+ : A+\operatorname{Neg}(A)\longrightarrow 0.\]

With the canonical association and unit maps understood, the corresponding snake equation is

\[\bigl(\varepsilon_A^++\operatorname{id}_A\bigr)\circ
\bigl(\operatorname{id}_A+\eta_A^+\bigr)=\operatorname{id}_A.\]

These are directed operations in the extended calculus. A forward passage through one part of a composite can encounter a turn that sends a negative value back through another part. The snake equation establishes the resulting round trip. The one-element type \lstinline!unit! is the multiplicative unit used later, while \(0\) supplies the additive boundary in these signatures.

We illustrate the computational role of directed turns with the three-state counter in §8 of Chen and Sabry\textquotesingle s paper. Start with a four-state counter on two bits, cycling through \lstinline!00!, \lstinline!01!, \lstinline!10! and \lstinline!11!. A negative connection routes the excluded \lstinline!11! state back through the increment circuit. The resulting positive computation is

\[00\longmapsto01,\qquad01\longmapsto10,\qquad10\longmapsto00.\]

The first two increments follow the underlying circuit directly. In the third, the intermediate \lstinline!11! takes the negative route and the next increment yields \lstinline!00!. The negative component therefore participates in computing the three-state cycle. A compiler may replace this construction with a simpler admitted realization once it establishes the same directed behavior.

Our interpretation uses the directed operations and coherence laws established by Chen and Sabry\textquotesingle s additive compact-closure result in Theorem 24. Their February 7, 2021 erratum corrects the further inverse-category assertion in Theorem 25, where distinct generalized reverses can satisfy the stated equations \hyperref[ref-chen-erratum]{{[}9{]}}. A library\textquotesingle s particular inverse is specified by its own round-trip equations.

\subsection*{2.2 Fractional Values and Value-Specific Elimination}\label{fractional-values-and-value-specific-elimination}
\addcontentsline{toc}{subsection}{2.2 Fractional Values and Value-Specific Elimination}

A fractional value participates in a different kind of computation. For a value \(v:A\), creation produces \(v\) together with a reciprocal value indexed by it. Elimination consumes the pair when the supplied value matches that index. Write \(r_v:1/v\) for the reciprocal and \(\star:I\) for the sole value of the unit type. The operations act as follows:

\[\eta_v^\times(\star)=(v,r_v),\qquad
\varepsilon_v^\times(w,r_v)=
\begin{cases}
\star,&w=v,\\
\mathsf{fail},&w\ne v.
\end{cases}\]

Our reference core keeps the reciprocal\textquotesingle s value-specific information in its type, following Chen and Sabry\textquotesingle s single-constructor representation. A Boolean created at \lstinline!true!, for example, can be negated twice and then eliminated with its reciprocal. After just one negation the carrier is still Boolean, but elimination reaches the mismatch state. The program\textquotesingle s intermediate computation determines whether elimination succeeds.

The reference machine treats mismatch as terminal, including in the combined negative/fractional calculus. A recoverable host interface needs an explicit rule for retaining the live pairing after an unsuccessful inspection. Our core interpretation preserves terminal mismatch. A library can expose a separate borrowed inspection, followed by consumption when the match succeeds, under its resource-retention law.

Value indexing also explains the categorical construction. A pointed type \((A,v)\) is a carrier with a distinguished value. Chen and Sabry establish compact closure for their pointed fractional construction in Theorem 36. Choosing one default value for an arbitrary unpointed carrier would change the relevant snake equation. In our interpretation the particular index therefore remains part of the program through creation, computation and elimination.

\subsection*{2.3 Composition and Search}\label{composition-and-search}
\addcontentsline{toc}{subsection}{2.3 Composition and Search}

We also examine how a context can determine an initially unknown value. In James and Sabry\textquotesingle s fractional calculus, creation introduces a fresh logic variable together with its reciprocal. Subsequent computation constrains that variable, and unification at the elimination site resolves the demand \hyperref[ref-james-dualities]{{[}1{]}}. James and Sabry\textquotesingle s satisfiability examples illustrate this computation through executable search.

Our value-indexed core starts with a specified value and follows its matching behavior. A logic-variable interface instead exposes the search and solution behavior of its calculus. A compiler may share representations or equality routines between the two, while retaining the source semantics that determine when a candidate is chosen and how failure proceeds. Search cost remains attached to the admitted search problem.

For both forms, an ordinary function can contain the operation or return a value that participates in it. The elaborator retains the additional use rules where such values cross a closure, branch or call. First-class composition thus includes a precise account of what can be copied, deferred and consumed.

\section*{3. Fractional Computation with Temporary Workspace}\label{fractional-computation-with-temporary-workspace}
\addcontentsline{toc}{section}{3. Fractional Computation with Temporary Workspace}

Temporary data frequently contributes to a result and then ceases to be needed. Fractional operations make the completion of that computation explicit. The program introduces the data at a particular value, uses it, and restores that value before eliminating the pair. Restoring the temporary data is often called \emph{uncomputation}. It permits a calculation to keep its useful output while completing the lifetime of its auxiliary state.

\subsection*{3.1 A Controlled-XOR Calculation}\label{a-controlled-xor-calculation}
\addcontentsline{toc}{subsection}{3.1 A Controlled-XOR Calculation}

Let \(W_w\) be the set of \(w\)-bit words, let \(x,y\in W_w\), and let \(h:W_w\to W_w\) be a pure total function. A controlled-XOR operation preserves its control and XORs a function of that control into its target. Applying the same operation twice restores the target, whatever the injectivity of \(h\).

Create an auxiliary word at the value \(0_w:W_w\), together with \(r_{0_w}:1/(0_w:W_w)\). The zero here is a word value. The auxiliary word supplies a temporary input to the controlled-XOR calculation and is restored before elimination:

{\def\LTcaptype{none} 
\begin{longtable}[]{@{}
  >{\raggedright\arraybackslash}p{(\linewidth - 4\tabcolsep) * \real{0.3333}}
  >{\raggedright\arraybackslash}p{(\linewidth - 4\tabcolsep) * \real{0.3333}}
  >{\raggedright\arraybackslash}p{(\linewidth - 4\tabcolsep) * \real{0.3333}}@{}}
\toprule\noalign{}
\begin{minipage}[b]{\linewidth}\raggedright
Operation
\end{minipage} & \begin{minipage}[b]{\linewidth}\raggedright
Data state
\end{minipage} & \begin{minipage}[b]{\linewidth}\raggedright
Fractional value
\end{minipage} \\
\midrule\noalign{}
\endhead
\bottomrule\noalign{}
\endlastfoot
Create auxiliary data & \((x,y,0_w)\) & \(r_{0_w}\) becomes available \\
XOR \(x\) into the auxiliary word & \((x,y,x)\) & Retained \\
XOR \(h(x)\) into the output & \((x,y\mathbin{\oplus}h(x),x)\) & Retained \\
XOR \(x\) into the auxiliary word again & \((x,y\mathbin{\oplus}h(x),0_w)\) & Ready for elimination \\
Eliminate the restored auxiliary pair & \((x,y\mathbin{\oplus}h(x))\) & Consumed \\
\end{longtable}
}

The reciprocal is part of the computation from creation to elimination. It allows the temporary word to be introduced into the composite and removed after it has contributed to the output. Its value index supplies the precise final-state requirement.

Using illustrative Clef library operations, the same construction is:

\begin{lstlisting}
let computeWithWorkspace h x y =
    let (a0, reciprocal) = introduce zeroWord
    let a1 = a0 ^^^ x
    let y1 = y ^^^ h a1
    let a2 = a1 ^^^ x
    eliminate (a2, reciprocal)
    (x, y1)
\end{lstlisting}

The word width is fixed by the enclosing interface. The reversible library interface sequences these operations in the displayed order, completing elimination before returning the result. Each update denotes an admitted controlled-XOR step in that fragment. The library\textquotesingle s implementation of \lstinline!h! carries its own workspace and cleanup contract. The example uses familiar binding syntax to expose the data dependencies, while the pairing operations contribute the value and usage judgments described in §4.

\textbf{Proposition 1 (workspace elimination).} Under these premises, elimination succeeds for every \((x,y)\in W_w^2\), and the resulting map \(F_h(x,y)=(x,y\mathbin{\oplus}h(x))\) is its own inverse.

\emph{Proof.} The auxiliary word after the second XOR with \(x\) is \((0_w\mathbin{\oplus}x)\mathbin{\oplus}x=0_w\), so it matches the creation index. Applying \(F_h\) twice XORs the unchanged word \(h(x)\) into \(y\) twice, giving \(y\). The control \(x\) is preserved throughout. \(\square\)

Omitting the restoration step leaves the auxiliary word equal to \(x\), so elimination succeeds exactly when \(x=0_w\). That difference is observable program behavior. A proof of restoration can discharge the equality statically and permit the runtime comparison to disappear. Storage removal is a further optimization justified by the realized dataflow and lifetime contract. Both transformations preserve the fractional operation\textquotesingle s source meaning.

\subsection*{3.2 Reciprocal Values at Library Boundaries}\label{reciprocal-values-at-library-boundaries}
\addcontentsline{toc}{subsection}{3.2 Reciprocal Values at Library Boundaries}

Our source notation \lstinline!Recip<A>! denotes a fractional carrier whose elaboration retains the particular value index. In mathematical judgments we write \(1/v\) to expose it. A library can infer the index from a constructor argument and keep it implicit in ordinary calls, just as dimensional parameters can remain implicit when the surrounding types determine them.

A fractional value can be returned, captured or passed to another operation when the corresponding resource use is admitted. Returning a pair exports the relationship needed for its eventual elimination. Capturing it in a one-shot thunk postpones that computation while retaining the index and its dependencies. A library that shares the immutable indexed data still tracks each consumable reciprocal separately.

This makes the application interface useful for ordinary workspace management as well as reversible programs. A computation can return a result together with a deferred completion operation. The completion executes the required restoration and elimination when demanded. Our memory analysis determines how long its captures remain available, and the fractional use rule determines how the pairing is completed.

The reciprocal\textquotesingle s value index and use rule can make runtime bookkeeping redundant. When its identity and use are statically determined, an unboxed realization can omit a separately stored reciprocal. Any required dynamic match and deferred restoration remain part of the computation. The typed construction and its evidence justify which information must be represented and which operations must execute.

\section*{4. Value Identity, Use and Inference}\label{value-identity-use-and-inference}
\addcontentsline{toc}{section}{4. Value Identity, Use and Inference}

The workspace example depends on two facts at elimination. The word must have returned to its initial value, and the reciprocal must belong to the live pairing being consumed. A type such as \lstinline!Word32! establishes the carrier. Our graph retains the more specific relationship needed to complete this operation.

\subsection*{4.1 A Finite Pairing Discipline}\label{a-finite-pairing-discipline}
\addcontentsline{toc}{subsection}{4.1 A Finite Pairing Discipline}

For the initial interpretation, take finite immutable data with pure decidable equality. The data contains no hidden consumable resources. We write \(\Gamma\) for the available ordinary values and \(\Delta\) for a finite map of live pairing instances. An entry \(\kappa\mapsto(A,v)\) identifies one instance, its carrier and its creation value. The symbol \(\kappa\) belongs to our host resource interpretation. Chen and Sabry\textquotesingle s value index remains \(v\).

The interface has the following operational state changes:

\[\begin{aligned}
\operatorname{introduce}(v):\quad
&\Delta\longmapsto\Delta\uplus\{\kappa\mapsto(A,v)\},\\
&\text{return }(v,r_{\kappa,v}),\quad\kappa\text{ fresh};\\[2pt]
\operatorname{eliminate}(w,r_{\kappa,v}):\quad
&\Delta\uplus\{\kappa\mapsto(A,v)\}\longmapsto\Delta,
\quad\operatorname{Eq}_A(w,v)=\mathsf{true}.
\end{aligned}\]

An authorized elimination with a false equality follows the reference mismatch behavior. Use of a consumed instance violates the resource interface. The static analysis represents these relationships through scoped resource variables and the admitted composition rules. A source operation executed repeatedly creates distinct dynamic instances, whose use is accounted for by the enclosing iteration or protocol invariant.

Equal creation values can belong to different instances. If two workspaces are both initialized to zero, restoring one establishes its value condition while its ownership identifies which pairing can be completed. Conversely, copying the immutable zero value supplies no additional reciprocal. This separation lets ordinary Clef sharing coexist with single consumption of each admitted resource.

Eliminating the reciprocal completes that pairing\textquotesingle s use. Releasing a physical allocation also requires the corresponding lifetime and alias conditions. Our DMM analysis supplies that storage account, including references that survive the local computation. A compiler can represent a pairing without an allocation when the surrounding program admits that realization.

\subsection*{4.2 Deferred and Shared Computation}\label{deferred-and-shared-computation}
\addcontentsline{toc}{subsection}{4.2 Deferred and Shared Computation}

A delayed operation carries the identities of the values it will use. Capturing an immutable snapshot retains the indexed value. Capturing a mutable location retains the location, with a separate requirement to establish which contents it contains when the operation runs. Our pairing discipline therefore connects a delayed elimination to its creation snapshot and to the intervening storage effects.

A memoized thunk presents a useful example. Its body may run once while several callers obtain the cached result. If that result contains a consumable reciprocal, the result interface must govern those later accesses. It can return a borrow for inspection or transfer ownership to one consumer. A justified duplication rule can support a richer interface when the selected resource calculus supplies it. Memoization and consumption have different points of use, and both remain visible in the graph.

The same rule applies when a continuation or closure is copied. A flat environment makes its captures explicit, allowing the compiler to check each captured resource under the closure\textquotesingle s use discipline. Our source operations retain these relationships through binding and generalization. An instance variable tied to a live resource remains rigid in its scope, so repeated instantiation cannot manufacture independently consumable copies.

\subsection*{4.3 Structural Proofs and Dimensional Inference}\label{structural-proofs-and-dimensional-inference}
\addcontentsline{toc}{subsection}{4.3 Structural Proofs and Dimensional Inference}

Our Tier 1 includes structural proofs for admitted negative and fractional constructions. A typing derivation supplies proof evidence in the associated logic, following Wadler\textquotesingle s propositions-as-types account \hyperref[ref-wadler2015]{{[}10{]}}. Inferring and composing their types establishes directed interface compatibility, retains value indices and checks permitted resource uses. The resulting derivations become premises for subsequent verification. An application-specific restoration equality or inverse law adds its local or library proof to this established structure.

An indexed resource can participate in the ordinary dimensional discipline. Kennedy\textquotesingle s units-of-measure system supplies principal inference for its specified integer-exponent measure fragment \hyperref[ref-kennedy1996]{{[}11{]}} \hyperref[ref-kennedy2009]{{[}12{]}}. F\# established that such information can be inferred in an ML-family programming language \hyperref[ref-syme]{{[}13{]}}. In our NTU, a force-valued input keeps its physical dimension when a negative operation redirects evaluation or a fractional operation pairs it with a reciprocal.

Three expressions make their respective roles explicit:

\begin{lstlisting}
Neg<float<N>>    // force data participating in directed evaluation
Recip<float<N>>  // a fractional value indexed by specified force data
float<N^-1>      // a numeric quantity with reciprocal-force dimension
\end{lstlisting}

The index is inferred or retained separately in the middle expression. A rational measure extension would solve linear equations over rational exponent vectors. Fractional value operations instead act on data and its use. Each contributes its own judgments to the same program.

Our geometric representation similarly retains parity, blade support and algebra signature according to their respective operations \hyperref[ref-phg]{{[}4{]}}. Range analysis provides enclosures, and memory analysis supplies lifetime conditions. The compiler combines established facts through declared transfer rules. The admitted source fragment determines the inference procedure and its cost. General value matching and reversible composition obtain their justification from equality procedures or applicable library proofs.

These judgments meet at the operation whose behavior they establish. In the controlled-XOR example, the word carrier and width follow from the ordinary operation types. The restoration equality follows from the XOR law. Resource analysis establishes the single elimination, and the lifetime analysis determines the storage that can be reused. A library interface can keep these derivations implicit while the compiler retains the premises needed to check and optimize each call.

\section*{5. Preservation Through Native Compilation}\label{preservation-through-native-compilation}
\addcontentsline{toc}{section}{5. Preservation Through Native Compilation}

Our compiler preserves these typed control and resource relationships through native compilation. The Program Semantic Graph associates each operation with its participating values, directed connections, resource instances and established premises. This keeps a restoration proof tied to the actual value compared or released and connects a negative turn to the control flow that realizes it. Within Clef Compiler Services, Baker owns that semantic construction. Alex witnesses its settled consequences in portable MLIR, and Composer\textquotesingle s backends realize the admitted operations for the selected target \hyperref[ref-fps]{{[}6{]}}.

A negative turn can therefore be realized through explicit control flow or a suitable continuation construction. A fractional creation, computation and elimination can be realized through ordinary values, storage and comparison operations. Their source meaning is retained in the correspondence between those operations and the graph. The middle end needs that correspondence even when the target has no dedicated negative or fractional type constructor.

\subsection*{5.1 Preserving a Fractional Match}\label{preserving-a-fractional-match}
\addcontentsline{toc}{subsection}{5.1 Preserving a Fractional Match}

Consider words in \(\{0,\ldots,255\}\) represented by their exact eight-bit encodings. Two words match precisely when their encodings match. Our source pairing also distinguishes the live instances to which those words belong. A local preservation result states the requirement independently of the chosen word width.

\textbf{Proposition 2 (matching under representation).} Let \(A\) be a finite data carrier with structural equality, let \(e:A\to B\) be an injective encoding with exact equality on encoded values, and let \(\rho\) be an injective renaming of live pairing identities. Translate \((\kappa,A,v)\) to \((\rho(\kappa),B,e(v))\) and a candidate \(w\) to \(e(w)\). The equality succeeds before translation exactly when it succeeds after translation. Distinct live instances remain distinct.

\emph{Proof.} Injectivity gives \(e(w)=e(v)\) if and only if \(w=v\). Injectivity of \(\rho\) preserves inequality between distinct instance identities. Applying both facts to the translated pairing establishes the claims. \(\square\)

The proposition justifies an exact encoded comparison. For an abstract equality relation, the encoding must preserve and reflect that relation instead. The corresponding native operation then receives the original match behavior, including its mismatch branch. A rounded conversion uses an explicit approximation relation and its own consumer contract.

The rest of the operation has further premises. The success branch consumes the admitted instance once, and any storage release respects surviving aliases. Constant folding a comparison uses its established equality while preserving those branch effects. In the workspace example, simplifying the restored word to zero can remove the comparison and the temporary storage when the admitted realization proves both transformations.

\subsection*{5.2 Composition and Parallel Use}\label{composition-and-parallel-use}
\addcontentsline{toc}{subsection}{5.2 Composition and Parallel Use}

Several independently completed workspaces can be released in a different order when their observable effects permit it. At the resource level, the relevant property is simple enough to state directly.

\textbf{Proposition 3 (independent successful elimination).} Suppose two live pairing instances have distinct identities, both candidate values match their indices, and their equality tests are pure. If the shared snapshots remain stable, eliminating either instance first produces the same final resource map.

\emph{Proof.} Each successful operation deletes its own entry from \(\Delta\) and leaves the other entry and \(\Gamma\) unchanged. Deletion of distinct entries commutes. \(\square\)

This provides a resource premise for parallel execution. The physical operations also require disjoint ownership or an applicable synchronization proof. General reordering retains the original observable behavior of errors and effects. A terminal mismatch, for example, can prevent a later operation from executing. Our compiler therefore uses the successful-elimination lemma in a context where its premises and the chosen observational equivalence hold.

Local graph rewrites need the same source correspondence. Lafont\textquotesingle s interaction nets and interaction combinators provide established accounts of local computation \hyperref[ref-lafont1990]{{[}14{]}} \hyperref[ref-lafont1997]{{[}15{]}}. An interpretation of negative and fractional type operations specifies which rewrites implement a directed turn or a value-indexed elimination. The resulting proof follows the source step and its operands through the rewrite, including any permitted information disposal.

\subsection*{5.3 Reusable Proofs and Mode Shifts}\label{reusable-proofs-and-mode-shifts}
\addcontentsline{toc}{subsection}{5.3 Reusable Proofs and Mode Shifts}

An application of the XOR law can justify many workspace eliminations. Our four verification tiers organize this reuse. Tier 1 supplies structural inference, including dimensions. Tier 2 generates local arithmetic and representation checks. Tier 3 instantiates reusable domain and system lemmas, including a library\textquotesingle s restoration theorem. Tier 4 expresses computational or probabilistic relations between executions \hyperref[ref-dbc]{{[}5{]}}.

The compiler generates the covered obligations and theorem applications at Tiers 1 through 3 without project-specific proof annotations. Library authors establish the general rules and their applicability conditions. Tier 4 admits project annotations that identify the intended relation, with the same library direction extending toward recognition of supported relational constructions from graph structure.

We use adjoint logic to organize reuse of established premises across the verification tiers. Hăvărneanu\textquotesingle s \emph{Classical Adjoint Logic} describes modes with explicit permissions for structural operations, including reuse or disposal of assumptions \hyperref[ref-havarneanu]{{[}16{]}}. The unit and counit of a mode adjunction govern proof transport. They are separate constructions from the computational turns and fractional operations defined in §2.

In our compiler account, a Tier 2 equality can serve directly as a premise of a Tier 4 relation when the admitted mode interface preserves its meaning and dependencies. A proof about a particular restored word remains tied to that word and its interpretation. Mode shifts organize this reuse without requiring a fresh intermediate proof at every tier. Kernel, solver and semantic-encoding dependencies remain attached to the actual evidence consumed.

These dependencies also define the scope of differential compilation. Editing a workspace transformation invalidates its restoration argument and the dependent lowering, while an independent segment retains its established results. FPS develops the segmented preservation mechanism \hyperref[ref-fps]{{[}6{]}}. NFT supplies the specific use and matching relationships that determine which dependent segments must be reconsidered.

\subsection*{5.4 A Negative Cost of Abstraction}\label{a-negative-cost-of-abstraction}
\addcontentsline{toc}{subsection}{5.4 A Negative Cost of Abstraction}

We use the recovery and resource relationships expressed by negative and fractional types to justify reductions in runtime work. This applies Stroustrup\textquotesingle s principle of negative-overhead abstraction, in which richer source structure gives an optimizer information it can use to reduce execution work \hyperref[ref-stroustrup-negative]{{[}17{]}}. In the workspace example, restoration evidence can remove a runtime comparison. Established identity and use can remove a separately stored reciprocal, and the lifetime argument can justify reuse of its auxiliary storage. The type-level account supplies the premises for eliminating work from a realization that would otherwise retain those checks and records.

The benefit begins with Tier 1 evidence established by negative and fractional type primitives and grows as local and domain proofs compose with them. A library can establish a general restoration or recovery rule, while the compiler checks its applicability to each use. Application code can therefore express the algorithm through those operations, with their invariants available throughout lowering. The following sections extend this account to replacing stored history with justified reconstruction, including the execution cost and error budget of that choice.

\section*{6. Reverse Computation on Demand}\label{reverse-computation-on-demand}
\addcontentsline{toc}{section}{6. Reverse Computation on Demand}

Our negative-type interface connects a directed computation to its permitted reversal as those operations are composed. A library supplies an inverse law on admitted states. A later recovery use determines which inputs, rule versions and resources must remain available. Where these premises justify reconstruction, the compiler can replace saved intermediate states with delayed reverse computation and preserve that justification through lowering. The exchange is especially useful when a sequence has a compact inverse recipe.

\subsection*{6.1 Exact State Reconstruction}\label{exact-state-reconstruction}
\addcontentsline{toc}{subsection}{6.1 Exact State Reconstruction}

Consider two \(w\)-bit words and a pure total function \(h:W_w\to W_w\). The transition

\[R(x,y)=(y,x\mathbin{\oplus}h(y))\]

has the inverse

\[R^{-1}(u,v)=(v\mathbin{\oplus}h(u),u).\]

This construction retains two words while each step changes which one supplies the control. It is useful as an elementary exact example because the operation that computes \(h\) may be many-to-one, while the surrounding transition retains the information required to reverse it.

\textbf{Proposition 4 (finite exact reconstruction).} For a fixed rule \(h\), both \(R^{-1}R\) and \(RR^{-1}\) are the identity on \(W_w^2\). Consequently, \(m\) inverse steps recover the input to \(m\) forward steps for every finite admitted count \(m\).

\emph{Proof.} Substituting \(R(x,y)\) into the inverse gives \(((x\mathbin{\oplus}h(y))\mathbin{\oplus}h(y),y)=(x,y)\). Substitution in the other order cancels \(h(u)\) in the same way. Repeated composition of the two identities proves the finite-step result. \(\square\)

For a fixed rule, working storage comprises the current pair, the workspace for \(h\) and a step counter. A changing rule schedule must be retained or reconstructible as well. The representation of a bounded counter has a fixed target width, while an unbounded mathematical count requires logarithmic storage in the count. The reconstruction uses this information in place of one saved pair per transition.

Fractional workspace operations can participate inside each step, as in §3. The inverse recipe determines how the state is recovered, and the fractional computation completes the use of its temporary data. Their composition supplies a concrete program with both directed and multiplicative resource structure.

\subsection*{6.2 Gradients Without a Stored Activation History}\label{gradients-without-a-stored-activation-history}
\addcontentsline{toc}{subsection}{6.2 Gradients Without a Stored Activation History}

Reverse differentiation needs the values at which local derivatives are evaluated. Conventional implementations retain many of those values in an activation tape. A reversible transition offers another source for them. Reconstructing the preceding state supplies the input for its derivative calculation when the backward pass reaches that step.

Let \(x_{t+1}=F_t(x_t;\theta)\) be differentiable and bijective on the admitted state domains, with inverse \(G_t\). For a terminal loss \(\ell(x_T)\), write \(\lambda_T=\nabla\ell(x_T)\). The backward calculation is

\[\begin{aligned}
x_t&=G_t(x_{t+1};\theta),\\
\lambda_t&=D_xF_t(x_t;\theta)^{\mathsf T}\lambda_{t+1},\\
\nabla_\theta\ell&=\sum_{t=0}^{T-1}
D_\theta F_t(x_t;\theta)^{\mathsf T}\lambda_{t+1}.
\end{aligned}\]

The last equation assumes an initial state independent of the shared parameters and a terminal loss with no direct parameter dependence. Other loss terms contribute their corresponding derivatives. The inverse reconstructs the primal state, while the transpose derivative transports the loss sensitivity. Both operations occur in the same backward calculation, with different mathematical roles.

Our learning application builds on the established exchange between activation storage and recomputation. For reversible residual networks, Gomez and colleagues combine reconstructed activations with ordinary backpropagation and discuss the error introduced by finite arithmetic \hyperref[ref-revnet]{{[}18{]}}. Maclaurin, Duvenaud and Adams provide a complementary finite-precision example in reversible learning, retaining information discarded by optimization steps so their reversal remains available \hyperref[ref-maclaurin]{{[}19{]}}. Our typed integration retains these reconstruction and arithmetic requirements with the operations to which they apply.

A Clef region can share an inverse recipe and a step cursor. Its deferred reverse calls retain the required parameter version and schedule, together with information needed to reproduce random choices or external observations. The graph tracks those captures and their lifetimes. Reconstructed states and derivative workspace then remain live for their required uses. This arrangement avoids retaining an activation at every step where reconstruction is admitted.

Fractional operations have an active role in the same calculation. They introduce local workspace, permit its use during recomputation and complete its elimination after the required restoration. A deferred completion must preserve the reciprocal and its value index across the delay. Our compiler can therefore reason jointly about the inverse computation, its transient storage and the resource operations that close each step.

A tape-free region has a specific storage account. Memoized thunks that retain all earlier states would preserve an activation history through their captures. We instead use the region\textquotesingle s shared recipe and retained reconstruction inputs to determine the necessary live state. Parameter storage, derivative workspace and any residual information remain included in that account.

\subsection*{6.3 Recovery Contracts for Ordinary Applications}\label{recovery-contracts-for-ordinary-applications}
\addcontentsline{toc}{subsection}{6.3 Recovery Contracts for Ordinary Applications}

Many application updates become reversible when their outputs retain sufficient recovery information. For a pure function \(f\), the embedding

\[x\longmapsto(f(x),x)\]

has a left inverse by projection on its image. A domain-specific inverse may permit a smaller retained value. The corresponding proof justifies each omission from the recovery record.

For example, a decision service can retain the input snapshot and rule version that produced its result:

\begin{lstlisting}
let evaluateForRecovery item rules =
    let inputs = readInputs item
    let result = evaluate rules inputs
    let recovery = (item, inputs, rules.version)
    (result, recovery)
\end{lstlisting}

The snapshot identifies the values actually used. Its rule version must identify an available implementation or retained rule state for recomputation. An interface can later replace this record with a checked inverse and a smaller remainder when the selected computation admits that reduction.

A reversible event region consequently has a declared recovery meaning. Restoring the exact state, recomputing a result within tolerance and issuing a compensating external action are different contracts. A fractional pairing can govern the particular recovery resource through creation and completion, while the associated domain law states what its reverse operation establishes. The application can use one library interface whose implementations preserve the selected guarantee.

\section*{7. Reconstruction Horizons and Arithmetic}\label{reconstruction-horizons-and-arithmetic}
\addcontentsline{toc}{section}{7. Reconstruction Horizons and Arithmetic}

An exact word permutation can be retraced for any finite admitted step count. Arithmetic over measured or learned quantities adds a different question. How far can an implementation retrace before accumulated error exceeds the application\textquotesingle s tolerance? A bound on that distance gives the compiler a reason to retain a checkpoint, select more accurate arithmetic or reconstruct on demand.

Our site\textquotesingle s \emph{The Lyapunov Window} design uses a tolerance to delimit this reconstruction span. The useful quantity for a program is a finite-time bound over its admitted state neighborhood. Lyapunov exponents describe the growth of perturbations along a dynamical evolution, while the compiler\textquotesingle s contract must cover the particular inverse operations and errors used in a reconstruction.

\subsection*{7.1 A Compositional Error Bound}\label{a-compositional-error-bound}
\addcontentsline{toc}{subsection}{7.1 A Compositional Error Bound}

Index steps by distance retraced from an anchor. Let \(x_j\) denote the states to be recovered, let \(\widehat x_j\) denote their reconstructions, and let \(E_j=\|\widehat x_j-x_j\|\) be their error. Suppose the realized reverse step \(\widehat G_j\) and an ideal map \(G_j\) satisfy

\[\begin{aligned}
\|G_j(u)-G_j(v)\|&\le L_j\|u-v\|,\\
\|\widehat G_j(\widehat x_j)-G_j(\widehat x_j)\|
+\|G_j(x_j)-x_{j+1}\|&\le\epsilon_j.
\end{aligned}\]

The bounds apply throughout the error neighborhood reached by the admitted segment. The second line includes both implementation error and any defect between the ideal reverse operation and the stated recovery target. If the target consists of already-rounded forward states, that defect includes information lost by the forward calculation.

\textbf{Proposition 5 (bounded reconstruction).} Under these premises,

\[E_{j+1}\le L_jE_j+\epsilon_j,\]

and after \(m\) reverse steps,

\[E_m\le E_0\prod_{j=0}^{m-1}L_j+
\sum_{i=0}^{m-1}\epsilon_i\prod_{j=i+1}^{m-1}L_j,\]

where an empty product equals one.

\emph{Proof.} Insert \(G_j(\widehat x_j)\) and \(G_j(x_j)\) between \(\widehat G_j(\widehat x_j)\) and \(x_{j+1}\). The triangle inequality and the two premises give the recurrence. Repeated substitution gives the displayed sum of amplified local errors. \(\square\)

For uniform bounds \(L_j\le L\) and \(\epsilon_j\le\epsilon\), this yields

\[E_m\le
\begin{cases}
L^mE_0+\epsilon\dfrac{L^m-1}{L-1},&L\ne1,\\[5pt]
E_0+m\epsilon,&L=1.
\end{cases}\]

With an exact anchor, \(L_j\le1\) and \(\epsilon_j\le2^{-20}\), a tolerance of \(2^{-12}\) admits 256 reverse steps. An inexact anchor uses part of the same budget through \(E_0\). For a history of \(N\) steps, anchors spaced \(m\) steps apart require storage proportional to \(N/m\), and a single recall requires at most \(m\) reverse steps, in addition to the recipe and its working state. The access pattern determines how often that recomputation is paid.

The bound describes the chosen state norm. In gradient computation, an error bound for the reconstructed primal also enters the derivative calculation\textquotesingle s sensitivity and arithmetic contract. Exact fractional elimination still uses the equality selected by its value index. A tolerance-based reconstruction can supply an approximate observation, while exact elimination requires a restored value or retained information sufficient to recover it exactly.

\subsection*{7.2 A Numerical Reversal Example}\label{a-numerical-reversal-example}
\addcontentsline{toc}{subsection}{7.2 A Numerical Reversal Example}

A symmetric step of a mechanical simulation makes the horizon concrete. Let \(\Phi_h\) be the mathematical fixed-step evolution and let \(S(q,p)=(q,-p)\) reverse momenta. For an admitted reversible integrator,

\[S\Phi_hS=\Phi_h^{-1}.\]

Starting from a state, take \(N\) forward steps, flip momenta, take another \(N\) steps with the same step size, then flip momenta again. The exact relation returns the initial state. A finite-arithmetic realization instead has a return residual governed by its actual local errors and sensitivity.

A reconstruction contract records the step rule and schedule along with the arithmetic assumptions. Changing a reduction order, adaptive timestep or state-dependent routing can change the realized map. The error account therefore belongs to the complete admitted execution, including those choices. A measured trajectory estimate can inform the selection of a policy, while a certified bound over the required neighborhood supplies its guarantee.

An adequate quire can accumulate represented products exactly before final rounding, reducing one contribution to local error. Division, square roots and rounded state updates retain their own contributions. Reducing those terms can extend the useful reconstruction span under unchanged dynamics. The resulting benefit is a later crossing of the declared tolerance, with its extent determined by the bound or experiment for that computation.

\subsection*{7.3 Bounds for a Realized Inverse Step}\label{bounds-for-a-realized-inverse-step}
\addcontentsline{toc}{subsection}{7.3 Bounds for a Realized Inverse Step}

Some reconstruction operations include a linear solve. A validated residual gives a direct route from that subroutine to the local error budget. For a nonsingular system \(Cx=d\), a computed \(\widehat x\), and compatible norms, suppose

\[K\ge\|C^{-1}\|,\qquad r\ge\|d-C\widehat x\|.\]

Then

\[\|\widehat x-x\|\le Kr,\]

because \(\widehat x-x=C^{-1}(C\widehat x-d)\). The residual bound includes error in its own evaluation. If the stored operands \(\widehat C,\widehat d\) approximate \(C,d\) within \(\eta_C,\eta_d\), respectively, a validated stored-system residual \(\widehat r\) gives

\[\|\widehat x-x\|\le
K\bigl(\widehat r+\eta_d+\eta_C\|\widehat x\|\bigr).\]

This expression supplies a local contribution to \(\epsilon_j\). It connects a mathematical inverse with the result returned by a particular arithmetic implementation.

We obtain a planning bound for elimination after Haar preconditioning from Theorem 1.2 in Urschel\textquotesingle s paper \hyperref[ref-urschel]{{[}20{]}}. For fixed nonsingular \(C\in\mathbb R^{n\times n}\) and an independent Haar-distributed orthogonal \(Q\), no-pivot elimination on \(QC\) satisfies

\[\Pr\!\left[\rho(QC)>T(n,\delta)\right]<\delta,
\qquad T(n,\delta)=\frac{14n^{3/2}\sqrt{\log(en)}}{\delta},
\quad0<\delta<1,\]

where, for the exact unit-lower/upper factorization \(B=LU\),

\[\rho(B)=\max\left\{\|L\|_{\max},
\frac{\|U\|_{\max}}{\|B\|_{\max}}\right\}.\]

If \(M\ge\|QC\|_{\max}\), the event \(\rho(QC)\le T\) gives \(\|L\|_{\max}\le T\) and \(\|U\|_{\max}\le TM\). A resulting coarse product bound is \((|L||U|)_{ij}\le nT^2M\). These are premises for range and arithmetic analysis of a covered elimination route. The transformed system \((QC)x=Qd\) preserves the exact solution and the two-norm condition number, so the separate sensitivity bound \(K\) still governs how residual error affects the recovered state.

Our compiler can use such a growth bound when selecting a candidate arithmetic realization, then bind the returned solution to its applicable residual or backward-error proof. Sampling and formation of the transformed operands require the contracts of the actual implementation. Deterministic guards protect storage and arithmetic coverage on the exceptional path. Once a realized solve meets its caller\textquotesingle s local error and risk contract, the enclosing reconstruction bound can be reused with another implementation meeting those premises. This permits arithmetic and solver choices to change while preserving the application\textquotesingle s recovery guarantee.

For a fixed reconstruction segment with conditional step-error risks bounded by \(\delta_j\), the probability that any step exceeds its admitted error is at most \(\sum_j\delta_j\). This union bound requires no independence between steps. Each conditional bound must apply to the input and sampling history at that step. The program can then select anchor spacing against both an error tolerance and a stated failure budget, while resource ownership remains an exact requirement.

\section*{8. Bidirectional State and Concurrent Recovery}\label{bidirectional-state-and-concurrent-recovery}
\addcontentsline{toc}{section}{8. Bidirectional State and Concurrent Recovery}

A computation can depend on information supplied later in a sequence. A document pass may need a suffix summary, and a service may wait for a reply before completing a resource use. These dependencies benefit from explicit connections between the value produced and the operation that will consume it. Their operational contracts determine how those connections can be evaluated or distributed.

\subsection*{8.1 A Two-Channel State Calculation}\label{a-two-channel-state-calculation}
\addcontentsline{toc}{subsection}{8.1 A Two-Channel State Calculation}

In our examination of forward and backward state channels, we draw lessons for their composition from Burton\textquotesingle s Tardis library \hyperref[ref-tardis]{{[}21{]}}. Suppressing the underlying monad, a step has the shape

\[C\,A=(B\times F)\longrightarrow(A\times B\times F),\]

where \(B\) is backward state and \(F\) is forward state. Binding computations \(m\) and \(k(a)\) connects their intermediate states through

\[\begin{aligned}
(a,b_0,f_1)&=m(b_1,f_0),\\
(r,b_1,f_2)&=k(a)(b_2,f_1).
\end{aligned}\]

The caller supplies \(b_2\) and \(f_0\), and the composite returns \(r,b_0,f_2\). The shared \(b_1\) and \(f_1\) connect the two steps in opposite directions. Burton\textquotesingle s implementation uses \lstinline!MonadFix! to establish these deferred connections.

For inputs \lstinline!2, 3, 5!, let each position receive a prefix sum from the left and a suffix sum from the right. It emits the two sums excluding its own value, adds its value to the forward sum, and adds it to the backward sum. With zero boundary values the outputs are \lstinline!(0, 8), (2, 5), (5, 0)!. Explicit forward and backward traversals realize this finite pure example. A library lemma can establish the correspondence with the source demand behavior.

Our graph retains the connected ports and the premises of that composition. Clef\textquotesingle s shared call-by-need values allow a connection to remain deferred until its consumer demands it. The library\textquotesingle s admitted productivity or scheduling rule establishes how the demand is supplied. An equation such as \(b=b\) describes a self-dependence without a value-producing definition, so it takes the library\textquotesingle s declared divergence or rejection behavior.

The state channels carry their own contents and operations. A channel using negative evaluation requires the corresponding directed interpretation, and one carrying a fractional value retains its value index and use rule. Lowering uses these relationships to produce an unboxed realization, preserving the required operations while erasing statically resolved type bookkeeping. These are substantive program constructions that can be composed with bidirectional state. Their library proofs establish the relevant identity and associativity laws under the chosen effects and equality.

\subsection*{8.2 In-Memory Event Recovery}\label{in-memory-event-recovery}
\addcontentsline{toc}{subsection}{8.2 In-Memory Event Recovery}

Event sourcing retains events so an application can replay the evolution of its state. Within a reversible region, checked inverse steps offer an additional implementation strategy. The program can keep a current state, a bounded set of anchors and the information needed to reconstruct the events between them. Exact audit requirements select exact reconstruction or retained evidence. A tolerance-based observation can instead use the numerical horizon of §7.

The static graph describes the operations and dependencies of the region. Each execution has its own event instances and values. A branch choice, overwritten value or external observation may contribute information that a particular inverse needs. Our resource interpretation associates that information with the event instance that owns it, so a storage optimization can justify which parts remain live.

A reversible event structure also records causal prerequisites for undoing an action. Ulidowski, Phillips and Yuen formalize event reversal with enabling and prevention conditions \hyperref[ref-events]{{[}22{]}}. For a local state update, the inverse may run after its dependent updates have been reversed. An external effect can require a compensating action with a different observable meaning. Our recovery interface retains the selected relation and protocol for each case.

Fractional operations govern the auxiliary computations inside that recovery process. A retained resource can pass through a worker, support a reconstructed calculation and return to its required value before elimination. The use condition remains attached to its actual instance across the sequence. This enables local inverse execution and resource completion to be optimized together, with causal and storage dependencies determining the permitted order.

\subsection*{8.3 Protocols Across Execution Boundaries}\label{protocols-across-execution-boundaries}
\addcontentsline{toc}{subsection}{8.3 Protocols Across Execution Boundaries}

A request/reply protocol illustrates why value identity and service order both matter. Suppose a worker receives \lstinline!(requestId, input)! and returns \lstinline!(requestId, result)!. The request identifier associates the reply with a dynamic invocation. The result contract establishes what was computed from that input. If the invocation transfers a consumable resource, the protocol also establishes which participant owns it before and after the reply.

We find that Qian, Kavvos and Birkedal\textquotesingle s client-server sessions provide a linear-logic account of client pools and shared servers \hyperref[ref-qkb]{{[}23{]}}. Their calculus admits different service orders, including examples with observable races. A pool\textquotesingle s well-formed composition therefore retains the protocol\textquotesingle s service-order behavior. A library seeking order-independent results supplies an additional law for its actual handler, such as commutation of the state updates under declared premises.

Our design applies the same principle to ordinary communicating components and to actor programs. An actor\textquotesingle s encapsulated state makes the protocol boundary explicit, while other programming styles can use the corresponding library interface. A session endpoint describes a protocol state. A fractional value describes the creation and elimination behavior defined by its calculus. Connecting the two requires a rule that preserves both the protocol transition and the resource\textquotesingle s value and use conditions.

Our DBC account draws on Iris for concurrent ownership reasoning, Actris for dependent separation protocols and Aneris for distributed systems proofs \hyperref[ref-dbc]{{[}5{]}} \hyperref[ref-iris]{{[}24{]}} \hyperref[ref-actris]{{[}25{]}} \hyperref[ref-aneris]{{[}26{]}}. Those libraries provide reusable reasoning for transferring and retaining resources across participants. An NFT library can instantiate the appropriate laws for its invocation and recovery protocol, with the premises bound to the graph\textquotesingle s actual participants.

BAREWire supplies the agreed representation at in-process, IPC and network boundaries. The payload includes any invocation or recovery fields required by the schema. Sending and receiving operations establish compatible interpretations and ownership transitions for those fields. Reliable delivery, duplicate acceptance and recovery after interruption are handled by the selected protocol. Their proofs determine when a remote computation has completed the resource use that a caller depends on.

\section*{9. Bayesian Computation and Retained Evidence}\label{bayesian-computation-and-retained-evidence}
\addcontentsline{toc}{section}{9. Bayesian Computation and Retained Evidence}

A learning system updates its knowledge when it receives an observation. The observation supplies concrete data to a statistical computation, whose prior and likelihood determine the result. A typed interface can preserve the observation\textquotesingle s physical dimension and provenance while the conditioning rule establishes its probabilistic meaning.

For finite parameter and observation spaces, write the prior as \(p(\theta)\) and the likelihood of observation \(d\) as \(L(d\mid\theta)\). When the marginal probability is positive,

\[Z(d)=\sum_\theta p(\theta)L(d\mid\theta)>0,
\qquad
p(\theta\mid d)=\frac{p(\theta)L(d\mid\theta)}{Z(d)}.\]

A pressure observation, for example, must be interpreted by a likelihood for the same physical quantity. Its range determines the admitted inputs, while its probability describes the weights assigned within the statistical account. Our graph retains the observation and likelihood used by the update. A zero marginal follows the library\textquotesingle s specified response for an observation excluded by that prior and likelihood.

\subsection*{9.1 A Reversible Finite Update}\label{a-reversible-finite-update}
\addcontentsline{toc}{subsection}{9.1 A Reversible Finite Update}

A fixed positive likelihood gives a concrete invertible calculation. Let \(\ell_\theta>0\) for every parameter value, and define \(T_\ell\) on the probability simplex by

\[T_\ell(p)_\theta=\frac{p_\theta\ell_\theta}
{\sum_\phi p_\phi\ell_\phi}.\]

\textbf{Proposition 6 (positive-likelihood reversal).} For a finite nonempty parameter space and a fixed strictly positive likelihood vector \(\ell\), the inverse of \(T_\ell\) is \(T_{1/\ell}\).

\emph{Proof.} Put \(q=T_\ell(p)\) and \(Z=\sum_\theta p_\theta\ell_\theta\). Then \(q_\theta/\ell_\theta=p_\theta/Z\), whose sum over \(\theta\) is \(1/Z\). Normalizing therefore recovers \(p\). Applying the same argument to \(1/\ell\) proves the other composition. \(\square\)

With prior \((1/2,1/2)\) and likelihood \((3/4,1/4)\), the normalizer is \(1/2\) and the posterior is \((3/4,1/4)\). Dividing by the retained likelihood gives \((1,1)\), which normalizes back to the prior. The arithmetic is exact over these rational values.

When the specified likelihood is already retained or can be compactly reproduced, this inverse permits the recovery record to omit a saved copy of the prior. The likelihood\textquotesingle s identity and availability become premises of the reverse operation, carried through composition to its point of use. A likelihood with zero components can discard prior mass, requiring additional recovery information for those components. Rounded implementations have the representation and error obligations of their selected arithmetic.

Fractional values participate through the computations that introduce and complete auxiliary state. A library may use an indexed workspace for the likelihood calculation, uncompute it after the useful output has been retained, and eliminate its reciprocal. It can similarly defer that completion until a reverse update has finished using the retained data. The numerical reciprocal \(1/\ell\) in Proposition 6 and the fractional value \(r_v:1/v\) have specified, separate roles within this program.

We use the weighted setting of Karwowski, Nielsen and Staton, §6 \hyperref[ref-karwowski-program-geometry]{{[}27{]}}, to make a further recovery example precise. Let \(X\) be a smooth state domain and \(w:X\to\mathbb R\) a smooth, finite, dimensionless logweight. Define \(F_w\) to execute \(\mathsf{score}(w(x))\) and return the unchanged \(x\). Its trace space has one element, its weight is \(e^{w(x)}\), and its logweight derivative and output derivative are \(Dw_x\) and \(\operatorname{id}_{T_xX}\). No random choice is introduced.

\textbf{Proposition 7 (weighted identity cancellation).} In exact real arithmetic, \(F_{-w}\circ F_w=\operatorname{id}_X=F_w\circ F_{-w}\) as weighted trace presentations under the canonical identification of singleton traces, including their derivatives. Their weighted metric transformers are inverses when the full logweight and value form is retained.

\emph{Proof.} The state remains \(x\) throughout, so the weights multiply to one and the logweight derivatives add to zero. To check the geometric statement, write \(a\in\mathbb R\) for a logweight tangent and \(\xi\in T_xX\) for a state tangent. The lifted derivative is the invertible shear

\[L_w(a,\xi)=(a+Dw_x[\xi],\xi),\qquad L_{-w}L_w=L_wL_{-w}=\operatorname{id}.\]

For a positive-semidefinite bilinear form \(B_x\) on \(\mathbb R\oplus T_xX\), our singleton transformer is

\[\mathbb A_w(B)_x=e^{w(x)}L_w^*B_x,\]

where \(L_w^*B_x(u,v)=B_x(L_wu,L_wv)\). Substitution gives \(\mathbb A_w\mathbb A_{-w}(B)=B\) and the reverse identity. In particular, starting with \(B_x((a,\xi),(b,\eta))=ab+h_x(\xi,\eta)\) gives

\[\mathbb A_w(B)_x((a,\xi),(b,\eta))
=e^{w(x)}\bigl[(a+Dw_x[\xi])(b+Dw_x[\eta])+h_x(\xi,\eta)\bigr].\]

The terms \(aDw_x[\eta]\) and \(bDw_x[\xi]\) participate in the inverse calculation. Projecting away these mixed terms between the two operations changes the transformer and invalidates that cancellation argument. \(\square\)

For a finite likelihood vector, the weight calculation specializes to multiplication by \(\ell_\theta\) followed by \(1/\ell_\theta\); Proposition 6 supplies the corresponding normalized-prior inverse. The recovery claim concerns this retained state and likelihood. It supplies no general rule for moving normalization across other program operations. Zero likelihoods, nonfinite logweights, and rounded implementations require their own admitted domains and recovery contracts.

A fractional implementation can create workspace at \(v\), compute the logweight while retaining \(x\), apply the two score operations using the same logweight, and uncompute the workspace to \(v\) before eliminating \((v,r_v)\). Proposition 7 establishes the ideal weighted cancellation; the workspace inverse establishes restoration, and the fractional rule checks its particular creation index and use. A source library must supply all three relationships. The identity does not derive the negative-type turns or fractional compact structure from metric pullback, and does not establish an unrestricted adjunction between sampling and conditioning. It instead gives a specific inverse law that our existing recovery interface can retain.

\subsection*{9.2 Evidence, Sampling and Reconstruction}\label{evidence-sampling-and-reconstruction}
\addcontentsline{toc}{subsection}{9.2 Evidence, Sampling and Reconstruction}

Retaining a posterior alone generally leaves the originating observation undetermined. For example, a likelihood independent of \(\theta\) leaves the prior unchanged for every observation with positive marginal probability. The observation\textquotesingle s identity or sufficient recovery information must therefore remain available when reconstruction is part of the interface.

We frame conditioning and retained information through stochastic kernels, using Fritz\textquotesingle s categorical account \hyperref[ref-fritz]{{[}28{]}}. In the finite case, composing kernels \(K(y\mid x)\) and \(M(z\mid y)\) gives \(\sum_y M(z\mid y)K(y\mid x)\). Conditioning forms a conditional distribution from a joint specification, while sampling produces a draw under the resulting distribution. Continuous conditioning uses its applicable measure-theoretic assumptions.

Our type and proof infrastructure retains these distinctions at the operations that consume them. A support proof identifies admitted values. A probability distribution weights those values, and a sample is a particular runtime occurrence. Fractional matching uses the specified equality and instance, including any retained random choices needed for reversal. An exact match can justify elimination even when the matched value originally came from a stochastic computation, provided its creation index and resource history have been preserved.

A parameter metric does not determine whether an execution can be reconstructed. For a counterexample, let a program draw a fair bit independently of every parameter and return a constant. Both trace probabilities are \(1/2\), their parameter scores are zero, and the constant output has zero velocity. Its score-\/-velocity form therefore vanishes, as does the form of the program with no draw. Forgetting the bit loses no geometry in these directions, yet the returned constant cannot identify the actual bit. A newly drawn fair bit has the same law and still need not equal the original occurrence. Restoring that occurrence requires retained information or an applicable reconstruction rule; zero geometric loss cannot discharge its fractional matching obligation.

Our graph consequently binds a geometric preservation claim to its trace law, output observation, parameter directions and program presentation, while binding exact recovery to the actual value and resource history. A transformation may satisfy either claim without satisfying the other. The natural-gradient application in \emph{Adaptive Domain Models} uses the former contract for an update metric and retains the latter wherever temporary sampled state must be restored.

A library can thus expose conditioning, sampling and recovery through ordinary Clef interfaces. The compiler applies the corresponding dimensional, use and probability laws to each construction. Where an application chooses an approximate posterior or reconstruction, its error relation remains part of the result consumed below that boundary.

\section*{10. Quantum Circuits and Auxiliary State}\label{quantum-circuits-and-auxiliary-state}
\addcontentsline{toc}{section}{10. Quantum Circuits and Auxiliary State}

A quantum circuit also introduces temporary resources, composes reversible operations and completes selected resource uses. Its mathematical interpretation adds complex amplitudes and a physical account of preparation and measurement. The classical workspace calculation in §3 offers a useful entry point. Computing into an auxiliary value and restoring it before release has a direct circuit analogue, with linearity extending the action from basis states to superpositions.

\subsection*{10.1 Compact Structure and Checked Gates}\label{compact-structure-and-checked-gates}
\addcontentsline{toc}{subsection}{10.1 Compact Structure and Checked Gates}

Finite-dimensional Hilbert spaces supply a dagger compact setting for quantum protocols. With a chosen orthonormal basis, a cup has the form \(\sum_i e_i^*\otimes e_i\), and its cap evaluates matching dual basis vectors. These maps satisfy the snake identities. Abramsky and Coecke develop their role in quantum protocols, and Selinger states the additional dagger structure and its compatibility laws \hyperref[ref-abramsky]{{[}29{]}} \hyperref[ref-selinger]{{[}30{]}}.

A compact dual supplies maps

\[\eta_A:I\longrightarrow A^*\otimes A,
\qquad\varepsilon_A:A\otimes A^*\longrightarrow I.\]

The snake identities describe the composite that returns a wire of type \(A\) or \(A^*\). The tensor product \(A\otimes A^*\) retains its own dimension. For a two-dimensional space it is four-dimensional, while \(I\) is one-dimensional. State preparation interprets the cup with its normalization and scalar factors. The displayed cup itself has norm \(\sqrt{\dim A}\).

A dagger assigns \(f^\dagger:B\to A\) to \(f:A\to B\). A unitary additionally satisfies

\[U^\dagger U=\operatorname{id}_A,
\qquad UU^\dagger=\operatorname{id}_B.\]

Our checked gate interface includes the specified operation, its adjoint and these inverse equations in the chosen ideal interpretation. The phase gate \(S=\operatorname{diag}(1,i)\) supplies a small example. Its adjoint is \(S^\dagger=\operatorname{diag}(1,-i)\), and both products equal the identity matrix. For checked \(U:A\to B\) and \(V:B\to C\), composition uses \(VU\) with adjoint \(U^\dagger V^\dagger\):

\[\begin{aligned}
(VU)^\dagger(VU)&=U^\dagger V^\dagger VU
=U^\dagger U=\operatorname{id}_A,\\
(VU)(VU)^\dagger&=VUU^\dagger V^\dagger
=VV^\dagger=\operatorname{id}_C.
\end{aligned}\]

Our checked gate interface makes these inverse laws available to circuit composition. A larger circuit can reuse them without reconstructing each gate\textquotesingle s inversion proof from its matrix representation. The compact operations, dagger and unitary laws each specify part of the interpretation used by the quantum library. Our negative and fractional source constructions enter that library through explicit interpretations with the corresponding preservation duties.

\subsection*{10.2 Circuits as First-Class Values}\label{circuits-as-first-class-values}
\addcontentsline{toc}{subsection}{10.2 Circuits as First-Class Values}

A closed circuit description is classical data with typed input and output ports. A host program can construct, return and compose such a value. Running it supplies live quantum resources at those ports. QWIRE establishes this separation between classical circuit values and quantum wires, while ReQWIRE develops verified reasoning about reversible circuits and auxiliary-qubit disposal \hyperref[ref-qwire]{{[}31{]}} \hyperref[ref-reqwire]{{[}32{]}}.

Our design connects these computational interfaces to the same graph representation used for other Clef programs. Fractional creation and elimination participate in the description of auxiliary-state computations. Directed reversal participates through the admitted reverse constructions. The graph retains the value relationships and use conditions that the chosen circuit interpretation requires.

A simulator interprets a circuit through classical representations of amplitudes or another admitted state description. A device interpretation maps it to physical operations with their own error account. The typed circuit can therefore support several realizations, each related to its declared semantics. Native CPU, GPU, NPU and FPGA routes provide the architectural setting for classical realizations, while a quantum-device route requires its corresponding preparation, gate and measurement contracts.

The computational cost follows the selected interpretation. A generic pure-state vector for \(n\) qubits has \(2^n\) complex amplitudes. A stabilizer description covers a more restricted family with different storage and execution costs. Our representation analysis uses such structural information when it is established for the particular circuit and state.

\subsection*{10.3 An Auxiliary Qubit Through Computation and Measurement}\label{an-auxiliary-qubit-through-computation-and-measurement}
\addcontentsline{toc}{subsection}{10.3 An Auxiliary Qubit Through Computation and Measurement}

Take two data qubits and an auxiliary qubit initialized to \(|0\rangle\). Two controlled-NOT operations compute parity into the auxiliary qubit:

\[|x_1x_2\rangle|0\rangle
\longmapsto|x_1x_2\rangle|x_1\mathbin{\oplus}x_2\rangle.\]

Before measurement, executing the controlled-NOTs in reverse order restores the auxiliary qubit to \(|0\rangle\) coherently. A larger reversible circuit can use parity to control a separate output while preserving the two data qubits. The same controlled-NOTs then uncompute the auxiliary qubit while retaining that output. The library\textquotesingle s fractional operations compose the creation, controlled use and restoration of that auxiliary state under its quantum resource interpretation. The completion rule keeps the condition for reuse attached to the circuit as that circuit is transformed.

Measuring the parity produces a different operation. For a data density matrix \(\rho\) and result \(m\in\{0,1\}\), the projectors, probabilities and conditional states are

\[\begin{aligned}
P_m&=\tfrac12\bigl(I+(-1)^mZ\otimes Z\bigr),\\
p_m&=\operatorname{Tr}(P_m\rho),\\
\rho_m&=P_m\rho P_m/p_m\qquad(p_m>0).
\end{aligned}\]

This is the usual measurement of two-qubit \(Z\) parity, within the stabilizer formalism described by Gottesman \hyperref[ref-gottesman]{{[}33{]}}. After the ideal measurement, the auxiliary qubit is in \(|m\rangle\). Applying \(X^m\) resets it to \(|0\rangle\), giving a state fact that the release operation can consume. The application retains the outcome and the conditional-state interpretation required by its subsequent work.

Coherent uncomputation and measurement followed by reset have different observable effects on the data. The latter selects a parity sector, retaining a classical outcome. Our graph therefore preserves which operation occurred, including its resource completion and probability law. A physical realization carries the preparation and measurement errors applicable to that interpretation.

Repeated measurements also have a resource cost. A fresh preparation, a reusable auxiliary qubit and a new copy of an input state are distinct resources. The protocol specifies which is required by the measurement task, and its quantitative analysis determines the preparation budget. Fractional operations express the relevant auxiliary computations once their state and use interpretation has been established.

\textbf{A shared preparation recipe.} Preparation identity can also affect the joint outcome law. Select a classical recipe for a normalized qubit \(|\psi\rangle=\psi_0|0\rangle+\psi_1|1\rangle\), and put \(p=|\psi_1|^2\). Uniform sampling on the real amplitude circle gives \(p\sim\operatorname{Beta}(\tfrac12,\tfrac12)\); Haar sampling of complex pure qubit states gives \(p\sim\operatorname{Beta}(1,1)\). These are the real and complex squared-coordinate laws described by Życzkowski and Sommers, §II.C and Appendix A \hyperref[ref-zyczkowski-sommers]{{[}34{]}}. Both ensembles have mean density matrix \(I/2\). A single computational-basis measurement therefore has the same averaged outcome law, \(\operatorname{Bernoulli}(1/2)\).

Now retain the selected recipe and use it for two fresh preparations. Under the ideal product-state preparation contract, the two basis outcomes \(Y_1,Y_2\) are independent conditional on that recipe, each with success probability \(p\). Their joint probability is

\[
\Pr(Y_1=Y_2=1)=\mathbb E[p^2]
=\begin{cases}
3/8,&\text{real-circle ensemble},\\
1/3,&\text{complex-Haar ensemble}.
\end{cases}
\]

The values follow from the second moments of the two Beta distributions. If the recipe is instead redrawn independently before each preparation, the same joint event has probability \((\mathbb E[p])^2=1/4\) for either ensemble. Preparing states from a retained classical description requires no cloning of an unknown state and does not reverse measurement.

Our graph contract would therefore distinguish the recipe identity from the identities of the two prepared resources, and retain the sampling and reuse policy whenever a consumer observes their joint outcomes. A replacement preserving the single-measurement law may violate this paired observation. This gives the preparation and resource distinctions above a concrete preservation obligation, with physical preparation errors supplied by the selected realization.

\subsection*{10.4 Geometric Representation and Arithmetic}\label{geometric-representation-and-arithmetic}
\addcontentsline{toc}{subsection}{10.4 Geometric Representation and Arithmetic}

A geometric representation can supply concrete implementations of particular operations. On the even subalgebra of \(\mathrm{Cl}(2,0)\), the bivector \(e_1e_2\) squares to \(-1\), and reversion sends it to \(-e_1e_2\). Thus \(a+b e_1e_2\) represents a complex scalar with reversion acting as conjugation. A selected matrix interpretation combines this conjugation with transpose to realize the adjoint.

More generally, an invertible versor supplies a checked conjugation action, and a normalized rotor has its inverse supplied by reversion. The orthogonal action on vectors uses the twisted adjoint, whose sign accounts for odd versors. Its outermorphism preserves homogeneous grade \hyperref[ref-lundholm]{{[}35{]}}. Our compiler retains the selected action and its algebra signature so that the corresponding inverse law applies to the operation being lowered.

An invertible representation must retain the quantities required by its reconstruction law. For example, normalizing a nonzero Euclidean vector while retaining its magnitude gives the representation

\[x\longmapsto(u,r)=\left(\frac{x}{\|x\|},\|x\|\right),
\qquad x=ru,\quad\|u\|=1,\ r>0.\]

The magnitude retains the information needed to reconstruct the original scale. Its physical dimension follows from the source quantity, while the unit direction is dimensionless. A fractional workspace computation can complete the temporary data used by this conversion after the required values have been retained. The graph distinguishes the magnitude needed for reconstruction from temporary values whose restored state permits reuse.

Amplitude representations need their own range and error account. Normalization bounds an amplitude\textquotesingle s magnitude by one, while a uniform state on \(2^n\) basis states has magnitudes \(2^{-n/2}\). Selecting a format therefore uses the actual amplitude range and accumulation requirements. Exact accumulation within an adequate quire addresses one part of that error account, with final rounding and subsequent operations retaining their own bounds.

BAREWire can carry classical circuit descriptions, control messages and simulator data through its admitted layouts. Live quantum resources use the physical protocol associated with the circuit interpretation. Our boundary contracts bind the classical control and observed results to that protocol, preserving the distinction between a circuit value and its execution.

\section*{11. A Checked Adiabatic Schedule}\label{a-checked-adiabatic-schedule}
\addcontentsline{toc}{section}{11. A Checked Adiabatic Schedule}

Quantum programs can also describe a continuous evolution rather than a sequence of gates. An adiabatic computation changes a Hamiltonian while seeking to keep the state near a selected eigenspace. This supplies a useful test of our design because its result depends on quantitative premises as well as the resource and operation structure.

For an ideal closed system with a self-adjoint Hamiltonian, Schrödinger evolution is unitary. A bound on departure from the instantaneous ground space additionally uses the path\textquotesingle s gap and regularity. A library can retain those premises with the schedule, so that selecting a duration becomes a checked application of the corresponding theorem.

\subsection*{11.1 A Two-Level Path with Explicit Bounds}\label{a-two-level-path-with-explicit-bounds}
\addcontentsline{toc}{subsection}{11.1 A Two-Level Path with Explicit Bounds}

Use dimensionless energy and time, the spectral norm, and the Pauli matrices \(X,Z\). Consider

\[H(s)=-(1-s)Z-sX,\qquad0\le s\le1.\]

The ground state begins at \(|0\rangle\) and ends at \(|+\rangle\). Direct diagonalization gives the gap and derivative bounds

\[g(s)=2\sqrt{(1-s)^2+s^2}\ge\sqrt2,
\qquad\|H'(s)\|=\sqrt2,
\qquad H''(s)=0.\]

We bound the departure probability for this rank-one ground space by \(A(1)^2\), using Theorem 3 of Jansen, Ruskai and Seiler \hyperref[ref-jrs]{{[}36{]}}. In this specialization its amplitude bound is

\[\begin{aligned}
A(1)\le\frac1\tau\biggl[&
\frac{\|H'(0)\|}{g(0)^2}+\frac{\|H'(1)\|}{g(1)^2}\\
&+\int_0^1\left(\frac{\|H''(s)\|}{g(s)^2}
+\frac{7\|H'(s)\|^2}{g(s)^3}\right)\,ds\biggr],
\end{aligned}\]

where \(\tau\) is the dimensionless duration and the initial state lies in the selected ground space. Using the endpoint gaps \(g(0)=g(1)=2\) and the lower bound on the gap in the integral yields

\[A(1)\le\frac1\tau\left(\frac{\sqrt2}{2}
+\frac{14}{(\sqrt2)^3}\right)=\frac{4\sqrt2}{\tau}.\]

The ideal departure probability is therefore at most \(\min(1,32/\tau^2)\). For a requested probability bound \(0<p<1\), the sufficient duration is \(\tau\ge4\sqrt{2/p}\). This is a conservative theorem-based choice for this path. For a physical Hamiltonian \(E_0H(t/T)\), the relation \(\tau=TE_0/\hbar\) restores the physical duration and its units.

The example supplies its gap and derivative bounds analytically. The compiler can check the instance against the theorem\textquotesingle s premises and the chosen probability budget. Preparation error and the error of the realized evolution contribute their additional bounds through the library\textquotesingle s composition rule.

\subsection*{11.2 Resources and Realization of the Schedule}\label{resources-and-realization-of-the-schedule}
\addcontentsline{toc}{subsection}{11.2 Resources and Realization of the Schedule}

A checked schedule carries the Hamiltonian path, duration and initial-state requirement. Its proof establishes the stated departure bound under the retained assumptions. The realization adds the interpretation of the chosen simulator or device operations. This gives the compiler concrete premises for representation, arithmetic and placement choices.

Negative and fractional computation can occur in the surrounding control and simulation program. A reverse operation follows a proved inverse of the admitted evolution, with its required inputs retained. Fractional operations introduce and eliminate the auxiliary values used to construct or execute the schedule. A quantum resource interface additionally accounts for preparation, measurement and the actual state relation used at release.

For a different Hamiltonian path, the same theorem application requires bounds over that path\textquotesingle s relevant interval. A finite set of sample points supports such a bound when accompanied by a covering argument. The reuse is in the theorem and its checked application, with the selected path supplying the mathematical premises.

\section*{12. Computational Structure and Engineering Choice}\label{computational-structure-and-engineering-choice}
\addcontentsline{toc}{section}{12. Computational Structure and Engineering Choice}

Negative and fractional types give recovery and resource completion a compositional place in the program\textquotesingle s type structure. Directed evaluation supports computations that retrace a path or route through a negative alternative. Fractional operations introduce auxiliary information, use it in a calculation and eliminate it under a value-specific rule. These forms make the algorithm\textquotesingle s control and resource relationships available to inference, verification and implementation as parts of the same program.

The negative cost of abstraction is concrete in these examples. A restoration proof can remove the workspace match, while statically determined identity and use can eliminate the reciprocal\textquotesingle s runtime payload. A reverse recipe with its required recovery data can replace a retained trajectory. For numeric regions, the reconstruction bound determines how far that exchange respects the requested accuracy and where anchors must be stored. Tier 1 supplies structural evidence shared by these operations, and local and domain proofs justify their particular reductions.

Those relationships also support changes during development. A revised computation invalidates the dependent matching, recovery or realization evidence, defining the segments that require renewed analysis and lowering. Unaffected segments retain their established proofs and realizations. Differential compilation therefore extends the benefit to the build loop, reusing both verified results and the lowered fragments that remain valid.

For an application developer, these constructions can appear as ordinary library operations for learning, event recovery or circuit composition. The library defines their computational laws, and our compiler applies the available inference and proof rules to the actual uses. Recovery and completion become compositional parts of the algorithm, with their invariants preserved into its native implementation.

\section*{Acknowledgments}\label{acknowledgments}
\addcontentsline{toc}{section}{Acknowledgments}

The author thanks Aram Hăvărneanu for development and contributions to adjoint classical logic and the mode-shift discipline that informs this work. The author also thanks Martin Coll for his work on interaction nets and ongoing engagement with the framework\textquotesingle s compiler strategy. Paul Snively\textquotesingle s contributions on delimited continuations, category and sheaf theory, and geometric algebra were pivotal to our investigation of the compact-closed extension. The author acknowledges John L. Gustafson\textquotesingle s posit arithmetic work as a foundation for our investigation of arithmetic precision and reconstruction horizons.

\section*{Software Availability}\label{software-availability}
\addcontentsline{toc}{section}{Software Availability}

The Clef language, Composer compiler, and supporting libraries described in this paper are developed under the Fidelity Framework project. Source repositories are available at \url{https://github.com/FidelityFramework}. The language specification, design rationale, and compiler documentation are published at \url{https://clef-lang.com}. Central components of the framework are dual-licensed; terms are detailed in each repository. The compiler design connects these components through the obligation and preservation interfaces described in this paper.

\section*{References}\label{references}
\addcontentsline{toc}{section}{References}

\leavevmode\protect\phantomsection\label{ref-james-dualities}{}{[}1{]} James, R. P., and Sabry, A. (2012). \emph{The Two Dualities of Computation: Negative and Fractional Types}. Technical Report, Indiana University, March 2012.

\leavevmode\protect\phantomsection\label{ref-chen}{}{[}2{]} Chen, C.-H., and Sabry, A. (2021). A computational interpretation of compact closed categories: Reversible programming with negative and fractional types. \emph{Proceedings of the ACM on Programming Languages}, 5(POPL), Article 9. \href{https://doi.org/10.1145/3434290}{DOI:10.1145/3434290}.

\leavevmode\protect\phantomsection\label{ref-dts}{}{[}3{]} Haynes, H. (2026). Dimensional type systems and deterministic memory management: Design-time semantic preservation in native compilation. \href{https://arxiv.org/abs/2603.16437}{arXiv:2603.16437}.

\leavevmode\protect\phantomsection\label{ref-phg}{}{[}4{]} Haynes, H. (2026). The program hypergraph: Multi-way relational structure for geometric algebra, spatial compute, and physics-aware compilation. \href{https://arxiv.org/abs/2603.17627}{arXiv:2603.17627}.

\leavevmode\protect\phantomsection\label{ref-dbc}{}{[}5{]} Haynes, H. (2026). Decidable by construction: Design-time verification for truly fearless systems. \href{https://arxiv.org/abs/2603.25414}{arXiv:2603.25414}.

\leavevmode\protect\phantomsection\label{ref-fps}{}{[}6{]} Haynes, H. (2026). Fixed-point scaffolding in the Clef programming language: Our theoretical grounding for type-preserving compilation and proof inference. \href{https://arxiv.org/abs/2606.02854}{arXiv:2606.02854}.

\leavevmode\protect\phantomsection\label{ref-adm}{}{[}7{]} Haynes, H. (2026). Adaptive domain models: Bayesian evolution, warm rotation, and principled training for geometric and neuromorphic AI. \href{https://arxiv.org/abs/2603.18104}{arXiv:2603.18104}.

\leavevmode\protect\phantomsection\label{ref-james-information}{}{[}8{]} James, R. P., and Sabry, A. (2012). Information effects. In \emph{POPL}, 73--84. \href{https://doi.org/10.1145/2103656.2103667}{DOI:10.1145/2103656.2103667}.

\leavevmode\protect\phantomsection\label{ref-chen-erratum}{}{[}9{]} Chen, C.-H., and Sabry, A. (2021). \emph{Errata of POPL 2021 paper: A Computational Interpretation of Compact Closed Categories: Reversible Programming with Negative and Fractional Types}. February 7. Correction to Theorem 25. \href{https://homes.luddy.indiana.edu/sabry/files/2021/02/errata.pdf}{Author version}.

\leavevmode\protect\phantomsection\label{ref-wadler2015}{}{[}10{]} Wadler, P. (2015). Propositions as types. \emph{Communications of the ACM}, 58(12), 75--84. \href{https://doi.org/10.1145/2699407}{DOI:10.1145/2699407}. \href{https://homepages.inf.ed.ac.uk/wadler/papers/propositions-as-types/propositions-as-types.pdf}{Author version}.

\leavevmode\protect\phantomsection\label{ref-kennedy1996}{}{[}11{]} Kennedy, A. (1996). \emph{Programming Languages and Dimensions}. PhD thesis, University of Cambridge, Technical Report 391. \href{https://www.cl.cam.ac.uk/techreports/UCAM-CL-TR-391.html}{Report}.

\leavevmode\protect\phantomsection\label{ref-kennedy2009}{}{[}12{]} Kennedy, A. (2009). Types for units-of-measure: Theory and practice. In \emph{Central European Functional Programming School}, LNCS 6299, 268--305. \href{https://doi.org/10.1007/978-3-642-17685-2_8}{DOI:10.1007/978-3-642-17685-2\_8}.

\leavevmode\protect\phantomsection\label{ref-syme}{}{[}13{]} Syme, D. (2020). The early history of F\#. \emph{Proceedings of the ACM on Programming Languages}, 4(HOPL), Article 75. \href{https://doi.org/10.1145/3386325}{DOI:10.1145/3386325}.

\leavevmode\protect\phantomsection\label{ref-lafont1990}{}{[}14{]} Lafont, Y. (1990). Interaction nets. In \emph{POPL}, 95--108. \href{https://doi.org/10.1145/96709.96718}{DOI:10.1145/96709.96718}.

\leavevmode\protect\phantomsection\label{ref-lafont1997}{}{[}15{]} Lafont, Y. (1997). Interaction combinators. \emph{Information and Computation}, 137(1), 69--101. \href{https://doi.org/10.1006/inco.1997.2643}{DOI:10.1006/inco.1997.2643}.

\leavevmode\protect\phantomsection\label{ref-havarneanu}{}{[}16{]} Hăvărneanu, A. (2026). \emph{Classical Adjoint Logic}. Pre-publication manuscript, July 12.

\leavevmode\protect\phantomsection\label{ref-stroustrup-negative}{}{[}17{]} Peterman, R. (2026, May 18). \href{https://www.developing.dev/p/creator-of-c-bell-labs-negative-overhead}{Creator of C++: Bell Labs, Negative Overhead Abstraction, Mistakes {|} Bjarne Stroustrup}. Interview with Bjarne Stroustrup. \emph{The Peterman Pod}. Transcript, 1:22:18--1:23:42.

\leavevmode\protect\phantomsection\label{ref-revnet}{}{[}18{]} Gomez, A. N., Ren, M., Urtasun, R., and Grosse, R. B. (2017). The reversible residual network: Backpropagation without storing activations. \href{https://arxiv.org/abs/1707.04585v1}{arXiv:1707.04585v1}.

\leavevmode\protect\phantomsection\label{ref-maclaurin}{}{[}19{]} Maclaurin, D., Duvenaud, D., and Adams, R. P. (2015). Gradient-based hyperparameter optimization through reversible learning. In \emph{ICML}, PMLR 37, 2113--2122. \href{https://proceedings.mlr.press/v37/maclaurin15.html}{Paper}.

\leavevmode\protect\phantomsection\label{ref-urschel}{}{[}20{]} Urschel, J. (2026). On the growth factor of random matrices. \href{https://arxiv.org/abs/2610.06785v1}{arXiv:2610.06785v1}.

\leavevmode\protect\phantomsection\label{ref-tardis}{}{[}21{]} Burton, D. \emph{Tardis: Bidirectional state monad transformer}. \href{https://github.com/DanBurton/tardis}{Source repository}, especially \lstinline!Control.Monad.Trans.Tardis!.

\leavevmode\protect\phantomsection\label{ref-events}{}{[}22{]} Ulidowski, I., Phillips, I., and Yuen, S. (2018). Reversing event structures. \emph{New Generation Computing}, 36, 281--306. \href{https://doi.org/10.1007/s00354-018-0040-8}{DOI:10.1007/s00354-018-0040-8}.

\leavevmode\protect\phantomsection\label{ref-qkb}{}{[}23{]} Qian, Z., Kavvos, G. A., and Birkedal, L. (2021). Client-server sessions in linear logic. \emph{Proceedings of the ACM on Programming Languages}, 5(ICFP), Article 62. \href{https://arxiv.org/abs/2010.13926}{arXiv:2010.13926}.

\leavevmode\protect\phantomsection\label{ref-iris}{}{[}24{]} Jung, R., Krebbers, R., Jourdan, J.-H., Bizjak, A., Birkedal, L., and Dreyer, D. (2018). Iris from the ground up: A modular foundation for higher-order concurrent separation logic. \emph{Journal of Functional Programming}, 28, e20. \href{https://www.mpi-sws.org/~dreyer/papers/iris-ground-up/paper.pdf}{Paper}.

\leavevmode\protect\phantomsection\label{ref-actris}{}{[}25{]} Hinrichsen, J. K., Krebbers, R., and Bengtson, J. (2022). Actris 2.0: Asynchronous session-type based reasoning in separation logic. \emph{Logical Methods in Computer Science}. \href{https://iris-project.org/actris/}{Paper and mechanization}.

\leavevmode\protect\phantomsection\label{ref-aneris}{}{[}26{]} Aneris project. \emph{Program logic for developing and verifying distributed systems}. \href{https://github.com/logsem/aneris}{Development and documentation}.

\leavevmode\protect\phantomsection\label{ref-karwowski-program-geometry}{}{[}27{]} Karwowski, J., Nielsen, F., and Staton, S. (2026). Compositional information geometry of probabilistic programs. \href{https://arxiv.org/abs/2610.17886v1}{arXiv:2610.17886v1}.

\leavevmode\protect\phantomsection\label{ref-fritz}{}{[}28{]} Fritz, T. (2020). A synthetic approach to Markov kernels, conditional independence and theorems on sufficient statistics. \emph{Advances in Mathematics}, 370, 107239. \href{https://arxiv.org/abs/1908.07021}{arXiv:1908.07021}.

\leavevmode\protect\phantomsection\label{ref-abramsky}{}{[}29{]} Abramsky, S., and Coecke, B. (2004). A categorical semantics of quantum protocols. In \emph{LICS}, 415--425. \href{https://doi.org/10.1109/LICS.2004.1319636}{DOI:10.1109/LICS.2004.1319636}.

\leavevmode\protect\phantomsection\label{ref-selinger}{}{[}30{]} Selinger, P. (2007). Dagger compact closed categories and completely positive maps. \emph{Electronic Notes in Theoretical Computer Science}, 170, 139--163. \href{https://doi.org/10.1016/j.entcs.2006.12.018}{DOI:10.1016/j.entcs.2006.12.018}.

\leavevmode\protect\phantomsection\label{ref-qwire}{}{[}31{]} Paykin, J., Rand, R., and Zdancewic, S. (2017). QWIRE: A core language for quantum circuits. In \emph{POPL}. \href{https://doi.org/10.1145/3009837.3009894}{DOI:10.1145/3009837.3009894}.

\leavevmode\protect\phantomsection\label{ref-reqwire}{}{[}32{]} Rand, R., Paykin, J., Lee, D.-H., and Zdancewic, S. (2019). ReQWIRE: Reasoning about reversible quantum circuits. \emph{EPTCS}, 287, 299--312. \href{https://arxiv.org/abs/1901.10118v1}{arXiv:1901.10118v1}.

\leavevmode\protect\phantomsection\label{ref-gottesman}{}{[}33{]} Gottesman, D. (1998). The Heisenberg representation of quantum computers. \href{https://arxiv.org/abs/quant-ph/9807006v1}{arXiv:quant-ph/9807006v1}.

\leavevmode\protect\phantomsection\label{ref-zyczkowski-sommers}{}{[}34{]} Życzkowski, K., and Sommers, H.-J. (2001). Induced measures in the space of mixed quantum states. \emph{Journal of Physics A: Mathematical and General}, 34, 7111--7125. \href{https://arxiv.org/abs/quant-ph/0012101v3}{arXiv:quant-ph/0012101v3}.

\leavevmode\protect\phantomsection\label{ref-lundholm}{}{[}35{]} Lundholm, D., and Svensson, L. (2016). \emph{Clifford Algebra, Geometric Algebra, and Applications}. Lecture notes, revised June 7. \href{https://www.math.lmu.de/~lundholm/clifford.pdf}{Author version}.

\leavevmode\protect\phantomsection\label{ref-jrs}{}{[}36{]} Jansen, S., Ruskai, M.-B., and Seiler, R. (2007). Bounds for the adiabatic approximation with applications to quantum computation. \emph{Journal of Mathematical Physics}, 48, 102111. \href{https://arxiv.org/abs/quant-ph/0603175v3}{arXiv:quant-ph/0603175v3}.
\end{document}